\documentclass[11pt]{article}
\usepackage{times}
\usepackage{eepic,epic}
\usepackage{latexsym}
\usepackage{amsfonts}
\usepackage{amsmath}
\usepackage{amssymb}
\usepackage{amsopn}
\usepackage{graphicx}
\makeatletter%
\def\nottoobig#1{{\hbox{$\left#1\vcenter to1.111\ht\strutbox{}\right.\n@space$}}}
\makeatother%

\makeatother%

\makeatletter%

\newcount\hour  \newcount\minutes  \hour=\time  \divide\hour by 60
\minutes=\hour  \multiply\minutes by -60  \advance\minutes by \time
\def\mmmddyyyy{\ifcase\month\or Jan\or Feb\or Mar\or Apr\or May\or Jun\or Jul\or
  Aug\or Sep\or Oct\or Nov\or Dec\fi \space\number\day, \number\year}
\def\hhmm{\ifnum\hour<10 0\fi\number\hour :%
  \ifnum\minutes<10 0\fi\number\minutes}
\def\Draft{{\it Draft of \mmmddyyyy}}

\def\ps@jtsheadings{%
\def\@oddhead{\it\rightmark\hfil\rm\thepage}%
\def\@oddfoot{\hfil\Draft}%
\if@twoside%
\def\@evenhead{\rm\thepage\hfil\it\leftmark}%
\def\@evenfoot{\Draft\hfil}%
\else
\let\@evenhead\@oddhead%
\let\@evenfoot\@oddfoot%
\fi%
}
\def\ps@jtsplain{%
\def\@oddhead{\hfil\Draft}%
\def\@oddfoot{\hfil\rm\thepage\hfil}%
\let\@evenfoot\@oddfoot%
\if@twoside \def\@evenhead{\Draft\hfil} \else \let\@evenhead\@oddhead \fi
}

\def\chaptermark#1{\markboth{\thechapter.\ #1}{\thechapter.\ #1}}%
\def\sectionmark#1{\markright{\thesection.\ #1}}

\def\section{\@startsection {section}{1}{\z@}
    {3.5ex plus1ex minus.2ex}{2.3ex plus.2ex}{\Large\bf}}
\def\subsection{\@startsection{subsection}{2}{\z@}
    {3.25ex plus1ex minus.2ex}{1.5ex plus.2ex}{\large\bf}}
\def\subsubsection{\@startsection{subsubsection}{3}{\z@}
    {3.25ex plus1ex minus.2ex}{1.5ex plus.2ex}{\normalsize\bf}}
\def\paragraph{\@startsection{paragraph}{4}{\z@}
    {3.25ex plus1ex minus.2ex}{1em}{\normalsize\bf}}
\def\subparagraph{\@startsection{subparagraph}{4}{\parindent}
    {3.25ex plus1ex minus.2ex}{1em}{\normalsize\bf}}

\makeatother%

\makeatletter \@beginparpenalty=10000 \makeatother

\def\underl#1 {\leavevmode\let\first=\relax\underli #1 }
\def\underli#1 {\ifx&#1\let\next=\relax\unskip
                \else\let\next=\underli\first\ulinebox{#1}\fi\let\first=\undersp\next}
\def\undersp{\penalty50\ulinebox{\space}\penalty50}
\def\ulinebox#1{\vtop{\hbox{\strut#1}\hrule}}%
\def\unice#1 {\underl #1 & }
\def\desclabel#1{\bf #1\hfil}
\def\desc{\list{}{%
\labelwidth=\leftmargin
\advance \labelwidth by -\labelsep
\let \makelabel=\desclabel}}

\makeatletter %

\newlength{\filength}
\newsavebox{\gcbox}
\sbox{\gcbox}{\framebox[\filength]{\rule{0ex}{2ex}}}

\newlength{\leftjustindent}
\newlength{\@leftjustindent}
\def\leftjust{\let\\\@leftjustcr\let\end\@endleftjust
  \addtolength{\@leftjustindent}{\leftjustindent}
  \vcenter\bgroup
  \halign\bgroup
    \hbox to\displaywidth{
      \rule{\@leftjustindent}{0ex}$\displaystyle##$\hfill
      }\crcr
}
\def\endleftjust{\crcr\egroup\egroup\endgroup}
\def\@endleftjust#1{\crcr\egroup\egroup\@checkend{#1}\endgroup}
\def\@leftjustcr{\crcr}

\newtheorem{theorem}{Theorem}%

\newcommand{\qedblob}{\mbox{\rule[-1.5pt]{5pt}{10.5pt}}}
\def\literalqed{{\ \nolinebreak\hfill\mbox{\qedblob\quad}}}

\def\qed{\literalqed}

\newtheorem{lemma}[theorem]{Lemma}

\newtheorem{proposition}[theorem]{Proposition}
\newcommand{\singlespacing}{\let\CS=
\@currsize\renewcommand{\baselinestretch}{1}\tiny\CS}
\newcommand{\singlespacingplus}{\let\CS=
\@currsize\renewcommand{\baselinestretch}{1.25}\tiny\CS}
\newcommand{\doublespacing}{\let\CS=
\@currsize\renewcommand{\baselinestretch}{1.75}\tiny\CS}
\newcommand{\draftspacing}{\let\CS=
\@currsize\renewcommand{\baselinestretch}{2.0}\tiny\CS}
\newcommand{\foospacing}{\let\CS=
\@currsize\renewcommand{\baselinestretch}{1.05}\tiny\CS}

\makeatother%

\mathcode`\0="0030      %
\mathcode`\1="0031
\mathcode`\2="0032
\mathcode`\3="0033
\mathcode`\4="0034
\mathcode`\5="0035
\mathcode`\6="0036
\mathcode`\7="0037
\mathcode`\8="0038
\mathcode`\9="0039

\newtheorem{definition}[theorem]{Definition}

\makeatletter
\clubpenalty=\@highpenalty
\widowpenalty=\@highpenalty
\makeatother

\makeatletter
\newcommand{\niceonespacing}{\let\CS=\@currsize\renewcommand{\baselinestretch}{1.1}\tiny\CS}\newcommand{\nicetwospacing}{\let\CS=\@currsize\renewcommand{\baselinestretch}{1.2}\tiny\CS}
\newcommand{\nicethreespacing}{\let\CS=\@currsize\renewcommand{\baselinestretch}{1.3}\tiny\CS}
\newcommand{\singlespacingplusplus}{\let\CS=\@currsize\renewcommand{\baselinestretch}{1.35}\tiny\CS}
\newcommand{\nicefourspacing}{\let\CS=\@currsize\renewcommand{\baselinestretch}{1.4}\tiny\CS}
\newcommand{\nicefivespacing}{\let\CS=\@currsize\renewcommand{\baselinestretch}{1.5}\tiny\CS}
\newcommand{\nicesixpacing}{\let\CS=\@currsize\renewcommand{\baselinestretch}{1.6}\tiny\CS}
\makeatother

\makeatletter
\def\@cite#1#2{[#1\if@tempswa , #2\fi]}
\makeatother

\makeatletter
\def\@citex[#1]#2{\if@filesw\immediate\write\@auxout{\string\citation{#2}}\fi
  \def\@citea{}\@cite{\@for\@citeb:=#2\do
    {\@citea\def\@citea{,\linebreak[0]}\@ifundefined
       {b@\@citeb}{{\bf ?}\@warning
       {Citation `\@citeb' on page \thepage \space undefined}}%
\hbox{\csname b@\@citeb\endcsname}}}{#1}}
\makeatother

\makeatletter
\def\ps@thesis{\def\@oddhead{\hfil\rm\thepage\hfil}\def\@oddfoot{}\def\@evenhead{\hfil\rm\thepage\hfil}\def\@evenfoot{}\def\chaptermark##1{}\def\sectionmark##1{}}
\makeatother

\makeatletter
\def\foobarpt{\textfont\z@\tenrm 
  \scriptfont\z@\ninrm \scriptscriptfont\z@\sevrm
\textfont\@ne\tenmi \scriptfont\@ne\ninmi \scriptscriptfont\@ne\sevmi
\textfont\tw@\tensy \scriptfont\tw@\ninsy \scriptscriptfont\tw@\sevsy
\textfont\thr@@\tenex \scriptfont\thr@@\tenex \scriptscriptfont\thr@@\tenex
\def\unboldmath{\everymath{}\everydisplay{}\@nomath\unboldmath
          \textfont\@ne\tenmi 
          \textfont\tw@\tensy \textfont\lyfam\tenly
          \@boldfalse}\@boldfalse
\def\boldmath{\@ifundefined{tenmib}{\global\font\tenmib\@mbi\@magscale1\global
        \font\tensyb\@mbsy \@magscale1\global\font
         \tenlyb\@lasyb\@magscale1\relax\@addfontinfo\@xiipt
              {\def\boldmath{\everymath
                {\mit}\everydisplay{\mit}\@prtct\@nomathbold
                \textfont\@ne\tenmib \textfont\tw@\tensyb 
                \textfont\lyfam\tenlyb\@prtct\@boldtrue}}}{}\@xiipt\boldmath}%
\def\prm{\fam\z@\tenrm}%
\def\pit{\fam\itfam\tenit}\textfont\itfam\tenit \scriptfont\itfam\ninit
   \scriptscriptfont\itfam\sevit
\def\psl{\fam\slfam\tensl}\textfont\slfam\tensl 
     \scriptfont\slfam\tensl \scriptscriptfont\slfam\tensl
\def\pbf{\fam\bffam\tenbf}\textfont\bffam\tenbf 
   \scriptfont\bffam\ninbf \scriptscriptfont\bffam\ninbf 
\def\ptt{\fam\ttfam\tentt}\textfont\ttfam\tentt
   \scriptfont\ttfam\nintt \scriptscriptfont\ttfam\nintt 
\def\psf{\fam\sffam\tensf}\textfont\sffam\tensf
    \scriptfont\sffam\tensf \scriptscriptfont\sffam\tensf
\def\psc{\@getfont\psc\scfam\@xiipt{\@mcsc\@magscale1}}%
\def\ly{\fam\lyfam\tenly}\textfont\lyfam\tenly 
   \scriptfont\lyfam\ninly \scriptscriptfont\lyfam\sevly
 \@setstrut \rm}

\makeatother

\newcommand{\p}{\mbox{\rm P}}

\newcommand{\DP}{\mbox{\rm DP}}
\newcommand{\np}{\mbox{\rm NP}}

\newcommand{\conp}{\mbox{\rm coNP}}

\newcommand{\sigmastar}{\mbox{$\Sigma^\ast$}}

\newcommand{\threecolor}{{{\tt 3\mbox{-}Colorability}}}

\newcommand{\threesat}{{{\tt 3\mbox{-}SAT}}}

\newcommand{\condition}{\,\nottoobig{|}\:}

\def\bhlevel#1{{{\mbox{\rm{}BH}_{#1}(\np)}}}
\newcommand{\bhnp}{{{\mbox{\rm{}BH}(\np)}}}
\def\exactcolor#1{{\tt{}Exact}\mbox{-}{#1}\mbox{-}{\tt{}Colorability}}

\newcommand\seq{\subseteq}

\newcommand\Lora{\, \Longrightarrow \ }

\newcommand{\naturalnumber}{\ensuremath{{  \mathbb{N} }}}
\def\nats{\naturalnumber}

\newenvironment{block}{\begin{list}{\hbox{}}{\leftmargin 1em
    \itemindent -1em \topsep 0pt \itemsep 0pt \partopsep 0pt}}{\end{list}}

\title{Exact Complexity of Exact-Four-Colorability}

\author{
J\"{o}rg Rothe\thanks{This work was
supported in part by grant NSF-INT-9815095/DAAD-315-PPP-g\"{u}-ab
and by a Hei\-sen\-berg Fellowship
of the Deut\-sche For\-schungs\-ge\-mein\-schaft.} \\ 
Abteilung f\"ur Informatik \\
Heinrich-Heine-Universit\"at D\"usseldorf \\
40225 D\"usseldorf, Germany \\
${\tt rothe@cs.uni\mbox{-}duesseldorf.de}$
}

\date{September 14, 2001}

\makeatletter
\def\@listI{\leftmargin\leftmargini \parsep 4.5pt plus 1pt minus 1pt\topsep
6pt plus 2pt minus 2pt \itemsep  2pt plus 2pt minus 1pt}

\let\@listi\@listI
\@listi
\makeatother

\begin{document}

\typeout{WARNING:  BADNESS used to suppress reporting!  Beware!!}
\hbadness=3000%
\vbadness=10000 %

\bibliographystyle{alpha}

\pagestyle{empty}
\setcounter{page}{1}

\sloppy

\pagestyle{empty}
\setcounter{footnote}{0}

\maketitle

\begin{abstract}
  Let $M_k \seq \nats$ be a given set that consists of $k$ noncontiguous
  integers.  Define $\exactcolor{M_k}$ to be the problem of determining
  whether~$\chi(G)$, the chromatic number of a given graph~$G$, equals
  one of the $k$ elements of the set $M_k$ exactly.  In 1987,
  Wagner~\cite{wag:j:min-max} proved that $\exactcolor{M_k}$ is
  $\bhlevel{2k}$-complete, where $M_k = \{6k+1, 6k+3, \ldots, 8k-1 \}$ and
  $\bhlevel{2k}$ is the $2k$th level of the boolean hierarchy over~$\np$.  In
  particular, for $k = 1$, it is DP-complete to determine whether $\chi(G) =
  7$, where $\DP = \bhlevel{2}$.  Wagner raised the question of how small the
  numbers in a $k$-element set $M_k$ can be chosen such that
  $\exactcolor{M_k}$ still is $\bhlevel{2k}$-complete.  In particular, for $k
  = 1$, he asked if it is DP-complete to determine whether $\chi(G) = 4$.
  In this note, we solve this question of Wagner and determine the precise
  threshold $t \in \{4, 5, 6, 7\}$ for which the problem $\exactcolor{\{t\}}$
  jumps from NP to DP-completeness: It is DP-complete to determine whether
  $\chi(G) = 4$, yet $\exactcolor{\{3\}}$ is in~$\np$.  More generally, for
  each $k \geq 1$, we show that $\exactcolor{M_k}$ is $\bhlevel{2k}$-complete
  for $M_k = \{3k+1, 3k+3, \ldots, 5k-1 \}$.
\end{abstract}

\setcounter{page}{1}
\pagestyle{plain}
\sloppy

\section{\boldmath{Exact-$M_k$-Colorability} and the Boolean Hierarchy over NP}

To classify the complexity of problems known to be NP-hard or coNP-hard, but
seemingly not contained in $\np \cup \conp$, Papadimitriou and
Yannakakis~\cite{pap-yan:j:dp} introduced~DP, the class of differences of two
NP problems.  They showed that DP contains various interesting types of
problems, including {\em uniqueness problems}, {\em critical graph problems},
and {\em exact optimization problems}.  For example, Cai and
Meyer~\cite{cai-mey:j:dp} proved the DP-completeness of
${\tt{}Minimal}\mbox{-}3\mbox{-}{\tt{}Uncolorability}$, a critical graph
problem that asks whether a given graph is not 3-colorable, but deleting any
of its vertices makes it 3-colorable.  A graph is said to be {\em
  $k$-colorable\/} if its vertices can be colored using no more than $k$
colors such that no two adjacent vertices receive the same color.  The {\em
  chromatic number of a graph~$G$}, denoted~$\chi(G)$, is defined to be the
smallest $k$ such that $G$ is $k$-colorable.

Generalizing~DP, Cai et
al.~\cite{cai-gun-har-hem-sew-wag-wec:j:bh1,cai-gun-har-hem-sew-wag-wec:j:bh2}
defined and studied the boolean hierarchy over~$\np$.  Their papers initiated
an intensive work and many papers on the boolean hierarchy;
e.g.,~\cite{wag:j:min-max,koe-sch-wag:j:diff,kad:joutdatedbychangkadin:bh,wag:j:bounded,bei:j:bounded-queries,cha:thesis,hem-rot:j:boolean}
to name just a few.
To define the boolean hierarchy, we use the symbols $\wedge$ and~$\vee$,
respectively, to denote the {\em complex intersection\/} and the {\em complex
  union\/} of set classes. That is, for classes $\mathcal{C}$ and
$\mathcal{D}$ of sets, define
\begin{eqnarray*}
\mathcal{C} \wedge \mathcal{D} & = &
\{ A \cap B \mid A \in \mathcal{C} \mbox{ and } B \in \mathcal{D}\}; \\ 
\mathcal{C} \vee \mathcal{D} & = &
\{ A \cup B \mid A \in \mathcal{C} \mbox{ and } B \in \mathcal{D}\}.
\end{eqnarray*}

\begin{definition} {\rm{}(\cite{cai-gun-har-hem-sew-wag-wec:j:bh1})}\quad
The {\em boolean hierarchy over $\np$\/} is inductively defined as follows:
\begin{eqnarray*}
\bhlevel{1} & = & \np, \\
\bhlevel{2} & = &  \np \wedge \conp, \\
\bhlevel{k} & = &  \bhlevel{k-2} \vee \bhlevel{2} 
                   \mbox{\quad for $k \geq 3$, and} \\
\bhnp & = &  \bigcup_{k \geq 1} \bhlevel{k}.
\end{eqnarray*}
\end{definition}

Equivalent definitions in terms of different boolean hierarchy normal forms
can be found in the
papers~\cite{cai-gun-har-hem-sew-wag-wec:j:bh1,wag:j:min-max,koe-sch-wag:j:diff};
for the boolean hierarchy over arbitrary set rings, we refer to the early work
by Hausdorff~\cite{hau:b:sets}.  Note that $\DP = \bhlevel{2}$.

In his seminal paper~\cite{wag:j:min-max}, Wagner provided sufficient
conditions to prove problems complete for the levels of the boolean hierarchy.
In particular, he established the following lemma for $\bhlevel{2k}$.

\begin{lemma}
\label{lem:klaus}
{\bf \cite[Thm.~5.1(3)]{wag:j:min-max}} \quad Let $A$ be some $\np$-complete
problem, let $B$ be an arbitrary problem, and let $k \geq 1$ be fixed.  

If there exists a polynomial-time computable function $f$ such that, for all
strings $x_1, x_2, \ldots, x_{2k} \in \sigmastar$ satisfying $(\forall j\,:\,
1 \leq j < 2k)\, [x_{j+1} \in A \Lora x_{j} \in A]$, it holds that
\begin{eqnarray}
\label{eq:klaus-codp}
\|\{i \condition x_i \in A\}\| \mbox{ is odd}
 & \Longleftrightarrow & f(x_1, x_2, \ldots, x_{2k}) \in B,
\end{eqnarray}
then $B$ is $\bhlevel{2k}$-hard.
\end{lemma}

For fixed $k \geq 1$, let $M_k = \{6k+1, 6k+3, \ldots, 8k-1 \}$, and define
the problem $\exactcolor{M_k} = \{ G \condition \chi(G) \in M_k \}$.  In
particular, Wagner applied Lemma~\ref{lem:klaus} to prove that, for each $k
\geq 1$, $\exactcolor{M_k}$ is $\bhlevel{2k}$-complete. For the special case
of $k = 1$, it follows that $\exactcolor{\{7\}}$ is DP-complete.

Wagner~\cite[p.~70]{wag:j:min-max} raised the question of how small the
numbers in a $k$-element set $M_k$ can be chosen such that $\exactcolor{M_k}$
still is $\bhlevel{2k}$-complete.  Consider the special case of $k = 1$.  It
is easy to see that $\exactcolor{\{3\}}$ is in NP and, thus, cannot be
DP-complete unless the boolean hierarchy collapses; see
Proposition~\ref{prop:exact-color} below.  Consequently, for $k = 1$, Wagner's
result leaves a gap in determining the precise threshold $t \in \{4, 5, 6,
7\}$ for which the problem $\exactcolor{\{t\}}$ jumps from NP to
DP-completeness.  Closing this gap, we show that it is DP-complete to
determine whether $\chi(G) = 4$.  More generally, answering Wagner's question
for each $k \geq 1$, we show that $\exactcolor{M_k}$ is
$\bhlevel{2k}$-complete for $M_k = \{3k+1, 3k+3, \ldots, 5k-1 \}$.

\section{Solving Wagner's Question}

\begin{proposition}
\label{prop:exact-color}
Fix any $k \geq 1$, and let $M_k$ be any set that contains $k$ noncontiguous
positive integers including~$3$.  Then, $\exactcolor{M_k}$ is in
$\bhlevel{2k-1}$; in particular, for $k = 1$, $\exactcolor{\{3\}}$ is
in~$\np$.  

Hence, $\exactcolor{M_k}$ is not $\bhlevel{2k}$-complete unless the boolean
hierarchy, and consequently the polynomial hierarchy, collapses.
\end{proposition}

\begin{proof}
  Fix any $k \geq 1$, and let $M_k$ be given as above.  
  Note that
\[
\exactcolor{M_k} = \bigcup_{i \in M_k} \exactcolor{\{i\}}.
\]
Since for each $i \in M_k$, $\exactcolor{\{i\}} = \{ G \condition \chi(G) \leq
i\} \cap \{ G \condition \chi(G) > i-1\}$ and since the set $\{ G \condition
\chi(G) \leq i\}$ is in~$\np$ and the set $\{ G \condition \chi(G) > i-1\}$ is
in~$\conp$, each of the $k-1$ sets $\exactcolor{\{i\}}$ with $i \in M_k -
\{3\}$ is in~$\DP$.  However, $\exactcolor{\{3\}}$ is even contained in~NP,
since it can be checked in polynomial time whether a given graph is
2-colorable, so $\{ G \condition \chi(G) > 2\}$ is in~$\p$.  It follows that
$\exactcolor{M_k}$ is in $\bhlevel{2k-1}$.~\qed
\end{proof}

\begin{theorem}
  For fixed $k \geq 1$, let $M_k = \{3k+1, 3k+3, \ldots, 5k-1 \}$.  Then,
  $\exactcolor{M_k}$ is $\bhlevel{2k}$-complete.  In particular, for $k = 1$,
  it follows that $\exactcolor{\{4\}}$ is $\DP$-complete.
\end{theorem}

\begin{proof}
  We apply Lemma~\ref{lem:klaus} with $A$ being the NP-complete problem
  $\threesat$ and $B$ being the problem $\exactcolor{M_k}$, where $M_k =
  \{3k+1, 3k+3, \ldots, 5k-1 \}$ for fixed~$k$.  The standard reduction
  $\sigma$ from $\threesat$ to $\threecolor$ has the following
  property~\cite{gar-joh:b:int}:
\begin{eqnarray}
\label{eq:three-four-color}
\phi \in \threesat \Lora \chi(\sigma(\phi)) = 3 & \mbox{ and } & 
\phi \not\in \threesat \Lora \chi(\sigma(\phi)) = 4.
\end{eqnarray}

Using the PCP theorem, Khanna, Linial, and
Safra~\cite{kha-lin-saf:j:fourcoloring-threecolorable-graphs} showed that it
is NP-hard to color a 3-colorable graph with only four colors.  Guruswami and
Khanna~\cite{gur-kha:c:fourcoloring-threecolorable-graphs} gave a novel proof
of the same result that does not rely on the PCP theorem.  We use their direct
transformation, call it~$\rho$, that consists of two subsequent
reductions---first from $\threesat$ to the independent set problem, and then
from the independent set problem to $\threecolor$---such that $\phi \in
\threesat$ implies $\chi(\rho(\phi)) = 3$, and $\phi \not\in \threesat$
implies $\chi(\rho(\phi)) \geq 5$.  Guruswami and
Khanna~\cite{gur-kha:c:fourcoloring-threecolorable-graphs} note that the graph
$H = \rho(\phi)$ they construct always is 6-colorable.  In fact, their
construction even gives that $H$ always is 5-colorable; hence, we have:
\begin{eqnarray}
\label{eq:three-five-color}
\phi \in \threesat \Lora \chi(\rho(\phi)) = 3 & \mbox{ and } & 
\phi \not\in \threesat \Lora \chi(\rho(\phi)) = 5.
\end{eqnarray}
To see why, look at the reduction
in~\cite{gur-kha:c:fourcoloring-threecolorable-graphs}.  The graph $H$
consists of tree-like structures whose vertices are replaced by $3 \times 3$
grids, which always can be colored with three colors, say~1, 2, and~3.  In
addition, some leafs of the tree-like structures are connected by leaf-level
gadgets of two types, the ``same row kind'' and the ``different row
kind.''~~The latter gadgets consist of two vertices connected to some grids,
and thus can always be colored with two additional colors.  The leaf-level
gadgets of the ``same row kind'' consist of a triangle whose vertices are
adjacent to two grid vertices each.  Hence, regardless of which 3-coloring is
used for the grids, one can always color one triangle vertex, say~$t_1$, with
a color $c \in \{1, 2, 3\}$ such that $c$ is different from the colors of the
two grid vertices adjacent to~$t_1$.  Using two additional colors for the
other two triangle vertices implies $\chi(H) \leq 5$, which proves
Equation~(\ref{eq:three-five-color}).

The join operation $\oplus$ on graphs is defined as follows: Given two
disjoint graphs $A = (V_A, E_A)$ and~$B = (V_B, E_B)$, their join $A \oplus B$
is the graph with vertex set $V_{A \oplus B} = V_A \cup V_B$ and edge set
$E_{A \oplus B} = E_A \cup E_B \cup \{ \{a, b\} \condition a \in V_A \mbox{
  and } b \in V_B\}$.  Note that $\oplus$ is an associative operation on
graphs and $\chi(A \oplus B) = \chi(A) + \chi(B)$.

Let $\phi_{1}, \phi_{2}, \ldots , \phi_{2k}$ be $2k$ given boolean formulas
satisfying $\phi_{j+1} \in \threesat \Lora \phi_{j} \in \threesat$ for each
$j$ with $1 \leq j < 2k$.  Define $2k$ graphs $H_{1}, H_{2}, \ldots , H_{2k}$
as follows.  For each $i$ with $1 \leq i \leq k$, define $H_{2i-1} =
\rho(\phi_{2i-1})$ and $H_{2i} = \sigma(\phi_{2i})$.  By
Equations~(\ref{eq:three-four-color}) and~(\ref{eq:three-five-color}),
\begin{eqnarray}
\label{eq:three-four-five}
\chi(H_j) & = & \left\{
\begin{array}{ll}
3 & \mbox{if $1 \leq j \leq 2k$ and $\phi_j \in \threesat$} \\
4 & \mbox{if $j = 2i$ for some $i \in \{1, 2, \ldots , k\}$ and $\phi_j
  \not\in \threesat$} \\
5 & \mbox{if $j = 2i - 1$ for some $i \in \{1, 2, \ldots , k\}$ and $\phi_j
  \not\in \threesat$.}
\end{array}
\right.
\end{eqnarray}
For each $i$ with $1 \leq i \leq k$, define the graph $G_i$ to be the disjoint
union of the graphs $H_{2i - 1}$ and~$H_{2i}$.  Thus, $\chi(G_i) =
\max\{\chi(H_{2i - 1}),\, \chi(H_{2i})\}$, for each $i$ with $1 \leq i \leq
k$.  The construction of our reduction $f$ is completed by defining
$f(\phi_{1}, \phi_{2}, \ldots , \phi_{2k}) = G$, where the graph $G =
\bigoplus_{i = 1}^{k} G_i$ is the join of the graphs $G_1, G_2, \ldots , G_k$.
Thus, 
\begin{equation}
\label{eq:sum-joins}
\chi(G) = \sum_{i = 1}^{k} \chi(G_i)
 = \sum_{i = 1}^{k} \max\{\chi(H_{2i - 1}),\, \chi(H_{2i})\}.  
\end{equation}
It follows from our construction that
\begin{eqnarray*}
\lefteqn{\|\{i \condition \phi_i \in \threesat\}\| \mbox{ is odd} } \\
 & \Longleftrightarrow & (\exists i \,:\, 1 \leq i \leq k)\, 
[\phi_{1}, \ldots , \phi_{2i - 1} \in \threesat \mbox{ and }
\phi_{2i}, \ldots , \phi_{2k} \not\in \threesat ] \\
 & \stackrel{(\ref{eq:three-four-five}),\, (\ref{eq:sum-joins})}
            {\Longleftrightarrow} & 
(\exists i \,:\, 1 \leq i \leq k)\, 
\left[\sum_{j = 1}^{k} \chi(G_j) = 3(i - 1) + 4 + 5(k - i)
                                 = 5k - 2i + 1 \right] \\
 & \stackrel{(\ref{eq:sum-joins})}
            {\Longleftrightarrow} & 
\chi(G) \in M_k = \{3k+1, 3k+3, \ldots, 5k-1 \} \\
 & \Longleftrightarrow & 
f(\phi_{1}, \phi_{2}, \ldots , \phi_{2k}) = G \in \exactcolor{M_k}.
\end{eqnarray*}

Hence, Equation~(\ref{eq:klaus-codp}) is satisfied.  By Lemma~\ref{lem:klaus},
$\exactcolor{M_k}$ is $\bhlevel{2k}$-complete.~\qed
\end{proof}

\bigskip

{\samepage
\noindent 
{\bf Acknowledgments.} \quad Interesting discussions with Klaus Wagner,
Venkatesan Guruswami, Edith and Lane Hemaspaandra, Dieter Kratsch, and Gerd
Wechsung are gratefully acknowledged.  }

{\small 

\bibliography{/sartre2/home2/rothe/BIGBIB/joergbib}

}

\end{document}